\documentclass[12pt,pra,showpacs,showkeys,amsmath,amssymb]{revtex4}

\begin{document}

\title{Nonlocal Dirac Equation for Accelerated Observers}

\author{Bahram Mashhoon}

\affiliation{Department of Physics and Astronomy\\
University of Missouri-Columbia\\
Columbia, Missouri 65211}

\begin{abstract}
The nonlocal theory of accelerated systems is extended to the propagation
of Dirac particles. The implications of nonlocality for the phenomenon
of spin-rotation coupling are discussed. The Lorentz-invariant nonlocal
Dirac equation is presented for certain special classes of accelerated
observers.
\end{abstract}



\maketitle

\section{Introduction\label{sec:1}}

The measurements of inertial observers are related to each other via
Lorentz transformations. All actual observers are accelerated, however.
To determine what an accelerated observer measures, a connection must
be established between the accelerated observer and the class of ideal
inertial observers. This is done in the standard theory of relativity
via the \textit{hypothesis of locality}, namely the assumption that
at each instant along its worldline an accelerated observer is equivalent
to an otherwise identical momentarily comoving inertial observer~\cite{1}.
Thus an accelerated observer passes through, or could be replaced
by, a continuous infinity of locally equivalent hypothetical momentarily
comoving inertial observers.

An inertial observer is endowed with an orthonormal tetrad frame;
therefore, it follows from the hypothesis of locality that an accelerated
observer in Minkowski spacetime carries an orthonormal tetrad $\lambda_{\;\;(\alpha)}^{\mu}(\tau)$
along its worldline. Here $\tau$ is the proper time of the observer.
The way in which the observer's tetrad frame is transported along
its worldline reveals its acceleration; that is,\begin{equation}
\frac{d\lambda_{\;\;(\alpha)}^{\mu}}{d\tau}=\Phi_{(\alpha)}^{\;\;\;\;(\beta)}\lambda_{\;\;(\beta)}^{\mu},\label{eq:1}\end{equation}
where $\Phi_{(\alpha)(\beta)}=-\Phi_{(\beta)(\alpha)}$ is the antisymmetric
\textit{acceleration tensor}. We note that $\lambda_{\;\;(0)}^{\mu}=dx^{\mu}/d\tau$
is the unit velocity vector that defines the local temporal axis of
the observer, while $\lambda^{\mu}_{\;\;(i)}$, $i=1,2,3$, are the unit
spacelike axes that define the observer's local spatial frame. The
``electric'' part of the acceleration tensor is given by $\Phi_{(0)(i)}=g_{(i)}$,
where $g^{(i)}(\tau)$ is the translational acceleration
of the observer with respect to the tetrad frame. Similarly, the {}``magnetic''
part of the acceleration tensor is given by the rotation of the observer's
local frame with respect to a locally nonrotating (i.e., Fermi-Walker
transported) frame; that is, $\Phi_{(i)(j)}=-\epsilon_{(i)(j)(k)}\Omega^{(k)}$,
where $\Omega^{(k)}(\tau)$ is the rotation frequency. Throughout
this paper, the signature of the metric is $-2$. One can naturally
define the magnitude of the translational acceleration $g(\tau)\geq0$
by $g^{2}(\tau)=-g_{(i)}g^{(i)}$ and that of the rotation frequency
$\Omega(\tau)\geq0$ by $\Omega^{2}(\tau)=-\Omega_{(k)}\Omega^{(k)}$.
It then proves useful to introduce the acceleration lengths $\mathcal{L}=c^{2}/g$
and $c/\Omega$ and acceleration times $\mathcal{L}/c=c/g$ and $1/\Omega$
to indicate the scales of spatial and temporal variation of the state
of the observer, respectively.

The hypothesis of locality originates from the mechanics of Newtonian
point particles. The state of such a particle is given by its position
and velocity; therefore, an accelerated particle and the corresponding
momentarily comoving inertial particle share the same state and are
thus pointwise equivalent. It follows that the description of accelerated
systems in Newtonian mechanics does not require any new hypothesis.
This was generalized to the relativistic mechanics of point particles
by Minkowski~\cite{2}. However, for radiation phenomena deviations
are expected from the hypothesis of locality that would be proportional
to $\lambda/\mathcal{L}$, where $\lambda$ is the wavelength of the
radiation.

The hypothesis of locality was introduced into the theory of relativity
by Lorentz as a useful approximation, and it was later essentially
adopted by Einstein~\cite{3}. As noted by Weyl~\cite{4}, the hypothesis
of locality amounts to an adiabatic approximation; that is, the length
and time scales associated with the phenomena under observation must
be negligibly small compared to the acceleration scales of the observer.
In fact, if all physical phenomena could be reduced to \textit{pointlike
coincidences} of particles and rays of radiation, then the hypothesis
of locality would be exactly valid. It would be interesting to go
beyond the hypothesis of locality and provide a general treatment
of the measurement problem for accelerated systems. This is the purpose
of the nonlocal theory of accelerated observers~\cite{5}, to which
we now turn.

Let $\psi$ be a basic radiation field in a global inertial frame
in Minkowski spacetime. For an accelerated observer, the measured
field along its worldline is $\widehat{\Psi}(\tau)$, while the field
as measured by the corresponding momentarily comoving inertial observers
is $\widehat{\psi}(\tau)=\Lambda(\tau)\psi(\tau)$, where $\Lambda$
is a matrix representation of the Lorentz group. The hypothesis of
locality implies that $\widehat{\Psi}(\tau)=\widehat{\psi}(\tau)$.
On the other hand, the most general linear relationship between $\widehat{\Psi}(\tau)$
and $\widehat{\psi}(\tau)$ that is consistent with causality may
be written as~\cite{6}\begin{equation}
\widehat{\Psi}(\tau)=\widehat{\psi}(\tau)+\int_{\tau_{0}}^{\tau}K(\tau,\tau')\widehat{\psi}(\tau')d\tau',\label{eq:2}\end{equation}
where $\tau_{0}$ is the initial instant at which the observer's acceleration
is turned on. Equation~\eqref{eq:2} deals with observable quantities
and is therefore manifestly invariant under the (global) Poincar\'e group;
moreover, the kernel $K$ must vanish in the absence of acceleration.
The postulated nonlocality of the field determination reflected in Eq.~\eqref{eq:2}
involves a weighted average over the past worldline of the accelerated
observer and is thus consistent with the ideas put forward by Bohr
and Rosenfeld~\cite{7}.

To determine the kernel $K$, a new physical postulate is needed;
therefore, we introduce the assumption that a basic radiation field
cannot stand completely still with respect to any (accelerated) observer~\cite{6}.
This is a direct generalization of a standard result of Lorentz invariance
for inertial observers to all observers; furthermore, it is naturally
consistent with the extension of Heisenberg's uncertainty principle
to accelerated systems.

To implement this new physical postulate, we take advantage of the
fact that for the Volterra integral Eq.~\eqref{eq:2} the relation
between $\widehat{\Psi}$ and $\psi$ is unique in the space of functions
of physical interest in accordance with the Volterra-Tricomi theorem~\cite{8}.
We therefore assume that a constant $\widehat{\Psi}$ must uniquely
correspond to a constant $\psi$, so that for any realistic variable
field $\psi$, $\widehat{\Psi}$ would never be a constant. In terms
of Eq.~\eqref{eq:2}, this means that $K$ and $\Lambda$ must be
related through\begin{equation}
\Lambda(\tau_{0})=\Lambda(\tau)+\int_{\tau_{0}}^{\tau}K(\tau,\tau')\Lambda(\tau')d\tau'.\label{eq:3}\end{equation}
It turns out that this integral equation does not provide a unique
solution for $K$ in terms of a given $\Lambda(\tau)$. A detailed investigation
reveals, however, that the only acceptable solution for $K$ is given
by~\cite{9,10}\begin{equation}
K(\tau,\tau')=k(\tau'),\label{eq:4}\end{equation}
where\begin{equation}
k(\tau')=-\frac{d\Lambda(\tau')}{d\tau'}\Lambda^{-1}(\tau').\label{eq:5}\end{equation}
The resulting nonlocal theory of accelerated systems is in agreement
with all available observational data. In fact, the nonlocal electrodynamics
of accelerated systems has been worked out in detail and the results
have been indirectly tested by the agreement between the consequences of
the nonlocal theory and the standard quantum mechanics of the interaction
of electromagnetic radiation with rotating electrons in the correspondence
limit~\cite{5}. Moreover, the nonlocal theory forbids the existence
of a fundamental scalar (or pseudoscalar) radiation field~\cite{6,11}.

Finally, it is important to remark that the nonlocality discussed
here is purely a vacuum effect and is not due to any limitations of
measuring devices; indeed, such limitations must be separately taken
into consideration.

This paper is concerned with the nonlocal aspects of Dirac's equation.
This equation is discussed from the standpoint of the hypothesis of
locality in the following section and the implications of nonlocality
for the phenomenon of spin-rotation coupling are considered in Section~\ref{sec:3}.
The nonlocal Dirac equation is presented in Sections~\ref{sec:4}
and \ref{sec:5}. Section~\ref{sec:6} contains a brief discussion
of our results. Some mathematical results regarding the resolvent
kernel are relegated to Appendix~\ref{sec:A}.

\section{Dirac spinors for accelerated observers\label{sec:2}}

Consider a class of fundamental inertial observers $\mathcal{F}$
at rest in a global inertial frame in Minkowski spacetime. According
to such an observer at $x^{\mu}=(ct,x,y,z)$, the Dirac spinor $\psi(x)$
satisfies the Dirac equation\begin{equation}
(i\hbar\gamma^{\alpha}\partial_{\alpha}-mc)\psi=0,\label{eq:6}\end{equation}
where $\gamma^{\alpha}$ are the usual Dirac matrices in the standard
representation~\cite{12}. Henceforth, we choose units such that
$\hbar=c=1$ in accordance with the standard convention~\cite{12}.
Under a Lorentz transformation $x\to x'$, where\begin{equation}
x^{\alpha}=L_{\;\;\beta}^{\alpha}x^{'\beta},\label{eq:7}\end{equation}
the Dirac equation takes the form~\cite{12}\begin{equation}
(i\gamma^{\alpha}\partial'_{\alpha}-m)\psi'(x')=0\label{eq:8}\end{equation}
such that the spinor transforms as\begin{equation}
\psi'(x')=\mathcal{S}(L)\psi(x),\label{eq:9}\end{equation}
where $\mathcal{S}(L)$ is the {}``spin transformation'' matrix
given by\begin{equation}
\mathcal{S}\gamma^{\alpha}\mathcal{S}^{-1}=L_{\;\;\beta}^{\alpha}\gamma^{\beta}.\label{eq:10}\end{equation}
This circumstance may be physically interpreted in terms of a different
class of inertial observers $\mathcal{F}'$ that are related to the
fundamental observers $\mathcal{F}$ by the Lorentz transformation~\eqref{eq:7}.
The form of Eq.~\eqref{eq:6} suggests that $\psi$ may be considered
a scalar under \textit{coordinate transformations}. The resulting
form-invariant Dirac equation would involve $\gamma^{'\alpha}$ given by \begin{equation}
\gamma^{\alpha}=L_{\;\;\beta}^{\alpha}\gamma^{'\beta}\label{eq:11}\end{equation}
and spinor $\tilde{\psi}'(x')=\psi(x)$ according to observers $\mathcal{F}'$.
To simplify matters, it would be useful to demand that observers $\mathcal{F}$ and
$\mathcal{F}'$ use the same representation of Dirac matrices. Under
a change of representation such that\begin{equation}
\mathcal{S}\gamma^{'\alpha}\mathcal{S}^{-1}=\gamma^{\alpha},\label{eq:12}\end{equation}
we have $\psi'(x')=\mathcal{S}\tilde{\psi}'(x')=\mathcal{S}\psi(x)$
and hence Eq.~\eqref{eq:9} is recovered; that is, the \textit{spin
transformation} is connected to a change of representation. Moreover, Eqs.~\eqref{eq:11}
and \eqref{eq:12} together imply Eq.~\eqref{eq:10}, which holds equally well for $\gamma^\alpha$
and $\gamma^{'\alpha}$, as can be easily checked. Thus in
dealing with spinors, we need to consider both coordinate and spin
transformations; however, under the former, the spinors remain invariant.

In a curved spacetime manifold, the generally covariant form of the
Dirac equation is given by~\cite{13}\begin{equation}
(i\gamma^{\mu}\nabla_{\mu}-m)\widehat{\psi}(x)=0,\label{eq:13}\end{equation}
where $\nabla_{\mu}=\partial_{\mu}+\Gamma_{\mu}$ and $\Gamma_{\mu}$
is the spin connection. The spinor $\widehat{\psi}$ is determined in
principle by a class of observers each with an orthonormal tetrad
$\lambda_{\;\;(\alpha)}^{\mu}$ such that\begin{equation}
g_{\mu\nu}(x)\lambda_{\;\;(\alpha)}^{\mu}\lambda_{\;\;(\beta)}^{\nu}=\eta_{(\alpha)(\beta)},\label{eq:14}\end{equation}
where $\eta_{(\alpha)(\beta)}$ is the Minkowski metric tensor. In
Eq.~\eqref{eq:13}, $\gamma^{\mu}$ is given by $\gamma^{\mu}=\lambda_{\;\;(\alpha)}^{\mu}\gamma^{(\alpha)}$,
where $\gamma^{(\alpha)}$ are the standard Dirac matrices in a local
inertial frame of the (Minkowski) tangent space. Moreover,
\begin{equation}
\Gamma_{\mu}=-\frac{i}{4}\lambda_{\nu(\alpha)}[\lambda_{\;\;(\beta)}^{\nu}]_{;\mu}\;\sigma^{(\alpha)(\beta)},\label{eq:15}\end{equation}
where\begin{equation}
\sigma^{(\alpha)(\beta)}=\frac{i}{2}[\gamma^{(\alpha)},\gamma^{(\beta)}].\label{eq:16}\end{equation}
This standard approach to the generally covariant Dirac equation is based on the hypothesis of locality and provides
the formalism by which a Dirac particle is minimally coupled to inertia
and gravitation. To avoid confusion, the standard Dirac matrices~\cite{12}
shall be henceforth denoted by $\gamma^{(\alpha)}$.

As discussed in detail in the previous section, we are interested
in a class of accelerated observers $\mathcal{A}$ in Minkowski spacetime;
therefore, we wish to use the covariant approach for $g_{\mu\nu}=\eta_{\mu\nu}$,
since we employ inertial Cartesian coordinates throughout this paper.
It then follows from Eq.\ \eqref{eq:15} that along the worldline
of an accelerated observer in Minkowski spacetime\begin{equation}
\Gamma_{\mu}\frac{dx^{\mu}}{d\tau}=\frac{i}{4}\left[\frac{d}{d\tau}\lambda_{\;\;(\alpha)}^{\nu}\right]\lambda_{\nu(\beta)}\;\sigma^{(\alpha)(\beta)}.\label{eq:17}\end{equation}
Equation~\eqref{eq:1} then implies that\begin{equation}
\Gamma_{\mu}\frac{dx^{\mu}}{d\tau}=\frac{i}{4}\Phi_{(\alpha)(\beta)}\;\sigma^{(\alpha)(\beta)}.\label{eq:18}\end{equation}
 The main purpose of this section is to determine $\widehat{\Psi}$; therefore, we must first find the connection between
spinor $\widehat{\psi}$ of the locally  inertial observers that are associated with $\mathcal{A}$ via the hypothesis of locality and the standard
Dirac spinor $\psi$ of the fundamental inertial observers $\mathcal{F}$.
Spinor $\psi$ satisfies the Dirac equation\begin{equation}
[i\gamma^{(\alpha)}\partial_{\alpha}-m]\psi(x)=0.\label{eq:19}\end{equation}
For $g_{\mu\nu}=\eta_{\mu\nu}$, Eqs.~\eqref{eq:13} and \eqref{eq:19}
refer to the same system of spacetime coordinates; therefore, let
us consider a spin transformation $\Lambda(x)$ such that\begin{equation}
\widehat{\psi}=\Lambda\psi;\label{eq:20}\end{equation}
then, Eq.~\eqref{eq:19} takes the form
\begin{equation}\label{eq:21} [i\gamma^\alpha (\partial_\alpha -\Lambda _{,\alpha }\Lambda^{-1})-m]\widehat{\psi}=0\end{equation}
once we choose $\Lambda (x)$ in such a way that
\begin{equation} \label{eq:22} \gamma^\alpha =\Lambda \gamma^{(\alpha )}\Lambda^{-1}=\lambda^\alpha_{\;\;(\beta)}\gamma^{(\beta)}.\end{equation}
In this case, a comparison of Eqs.~\eqref{eq:13} and \eqref{eq:21} indicates that
\begin{equation} \label{eq:23} \Gamma_\alpha =-\Lambda_{,\alpha}\Lambda^{-1}.\end{equation}
Thus along the worldline of an accelerated observer,
\begin{equation}\label{eq:24} \Gamma_\alpha \frac{dx^\alpha}{d\tau }=-\frac{d\Lambda}{d\tau}\Lambda^{-1}.\end{equation}
Therefore, Eq.~\eqref{eq:5} implies that the kernel of our nonlocal ansatz is given by
\begin{equation}\label{eq:25} k(\tau )=\Gamma_\alpha \frac{dx^\alpha}{d\tau }.\end{equation}
We finally have the result that $\widehat{\psi}=\Lambda\psi$ with
\begin{equation}\label{eq:26} \Lambda (\tau )=e^{-\int^\tau _{\tau_0}k(\tau ')d\tau '}\Lambda (\tau _0),\end{equation}
where the kernel $k$ is given by Eq.~\eqref{eq:18},
\begin{equation}\label{eq:27} k(\tau )=\frac{i}{4} \Phi _{(\alpha )(\beta )}(\tau )\;\sigma ^{(\alpha )(\beta )},\end{equation}
and $\widehat{\Psi} $ is then determined by our nonlocal ansatz~\eqref{eq:2}.

A remark is in order here regarding the close analogy between $\Lambda$ in Eq.~\eqref{eq:22} and $\mathcal{S}$ in the
corresponding Eq.~\eqref{eq:10}. It is clear from this correspondence that $\Lambda$ must be
interpreted as a pointwise spin transformation related to a local Lorentz transformation in accordance
with the hypothesis of locality. That is, $dx^\alpha=L^\alpha_{\;\;\beta} dx^{'\beta}$, derived from Eq.~\eqref{eq:7},
is the exact analog of the pointwise Lorentz transformation to the local frame of the accelerated observer:
$dx^\alpha =\lambda^\alpha_{\;\;(\beta )} dx^{(\beta )}$.Thus for any fixed $\tau \geq \tau_0$, Eq.~\eqref{eq:26}
resembles what one would find for the spin transformation under the Lorentz group~\cite{12}; the significance of
this resemblance is that it is simply the expression of the hypothesis of locality for a Dirac particle.

Now that we have general expressions for the spin transformation $\Lambda$ and the kernel $k$, it is interesting to
illustrate some of the consequences of the nonlocal theory in connection with the phenomenon of spin-rotation coupling
for a Dirac particle. It should be mentioned in this connection that Dirac's equation in accelerated systems has been
investigated by a number of authors (see Refs.~\cite{13}-\cite{16} and references therein) and the inertial properties
of a Dirac particle have been elucidated; for a more complete discussion of the background material and list of references,
see Ref.~\cite{17}. Spin-rotation coupling and its observational aspects have been treated in Refs.~\cite{18}-\cite{26}.

\section{Spin-rotation coupling\label{sec:3}}
Consider a class of observers rotating uniformly with frequency $\Omega$ about the $z$ axis. To characterize the worldline
of these observers, let us assume that for $t<0$, a typical observer moves uniformly parallel to the $y$ axis with
$x=r$, $y=r\Omega t$ and $z=z_0$, where $r$, $\Omega$ and $z_0$ are constants with $r\geq 0$, $\Omega \geq 0$ and
$-\infty <z_0<\infty$. At $t=0$, the observer follows a circle of radius $r$ with $x=r\cos \phi$, $y=r\sin \phi$ and $z=z_0$,
where $\phi =\Omega t=\gamma \Omega \tau$. Here $\gamma$ is the observer's Lorentz factor that corresponds to $\beta =r\Omega$.
The observer's orthonormal tetrad frame for $t\geq 0$ is thus given by
\begin{align}\label{eq:28} \lambda^\mu_{\;\;(0)} &=\gamma (1,-\beta \sin \phi ,\beta \cos \phi ,0),\\
\label{eq:29} \lambda^\mu _{\;\;(1)}&=(0,\cos \phi ,\sin \phi ,0),\\
\label{eq:30} \lambda^\mu _{\;\;(2)}&=\gamma (\beta ,-\sin \phi ,\cos \phi ,0),\\
\label{eq:31} \lambda^\mu _{\;\;(3)}&=(0,0,0,1).\end{align}
It is now possible to calculate $\Lambda $ and $k$ in this case using Eqs.~\eqref{eq:26} and \eqref{eq:27}.

The nonzero components of the acceleration tensor can be determined from the observer's centripetal acceleration $\Phi_{(0)}^{\;\;\;\;\;\;(1)} =-\beta \gamma^2\Omega$ and its rotational frequency $\Phi _{(1)}^{\;\;\;\;\;\;(2)} =\gamma^2\Omega$. Thus the kernel $k$, given by Eq.~\eqref{eq:27}, is a constant matrix in this case given by
\begin{equation}\label{eq:32} k=-\frac{i}{2}\gamma^2\Omega \begin{bmatrix} \sigma^3 & -i\beta \sigma^1\\ -i\beta \sigma^1 & \sigma^3\end{bmatrix} ,\end{equation}
where $\sigma^i$, $i=1,2,3$, is a Pauli matrix~\cite{12}. Using the fact that $-4k^2/(\gamma\Omega)^2$ is the identity matrix, we find that
\begin{gather}\label{eq:33} e^{-k\tau }=\begin{bmatrix} A& B\\ B & A\end{bmatrix},\\
\label{eq:34}A=\cos \frac{\phi}{2}\; I+i\gamma \sin\frac{\phi}{2}\;\sigma^3,\quad B=\beta \gamma\sin \frac{\phi}{2}\;\sigma^1,\end{gather}
where $I$ is the $2\times 2$ unit matrix. It remains to determine $\Lambda (\tau _0)=\Lambda _0$, which simply corresponds to a boost in the $y$ direction with speed $\beta $, i.e.
\begin{equation}\label{eq:35} \Lambda_0 =\begin{bmatrix} a & b \\ b & a\end{bmatrix} ,\quad a=\sqrt{\frac{1}{2}(\gamma +1)}\; I,\quad b=-\sqrt{\frac{1}{2}(\gamma -1)}\;\sigma^2.\end{equation}

We now imagine positive-energy plane-wave solutions of the free Dirac equation propagating along the $z$ axis with momentum $p$ and spin parallel $(\psi _+)$ or antiparallel $(\psi _-)$ to the $z$ axis. Therefore,
\begin{align}\label{eq:36} \psi_\pm &=w_\pm e^{-iEt+ipz},\\
w_+ &=N\begin{bmatrix} 1\\ 0\\ \rho \\0\end{bmatrix} ,\quad w_-=N\begin{bmatrix} 0 \\ 1\\ 0\\-\rho \end{bmatrix},\end{align}
where $\rho =p/(m+E)$, $E=\sqrt{m^2+p^2}$ and $m$ is the mass of the Dirac particle. Here $N$ is a certain positive normalization factor; its exact value is unimportant for our present considerations. The Dirac spinor according to the hypothesis of locality for the uniformly rotating observer with parameters $(r,\Omega ,z_0)$ is given by
\begin{align}\label{eq:38} \widehat{\psi}_\pm &=\widehat{w}_\pm e^{-iE'_\pm \tau +ipz_0},\\
\label{eq:39} \widehat{w}_+ &=N'\begin{bmatrix} \gamma +1\\ -i\beta \gamma \rho \\ (\gamma+1)\rho\\ -i\beta \gamma\end{bmatrix} ,\quad \widehat{w}_- =N'\begin{bmatrix} -i\beta \gamma \rho \\ \gamma+1\\ i\beta \gamma\\ -(\gamma +1)\rho \end{bmatrix},\end{align}
where $N'=N\sqrt{2(\gamma +1)}$ and
\begin{equation} \label{eq:40} E'_\pm =\gamma \left( E\mp \frac{1}{2}\Omega \right).\end{equation}
Let us digress here and mention that a detailed investigation reveals that Eq.~\eqref{eq:40} is valid for negative-energy Dirac spinors as well.

Equation~\eqref{eq:40} clearly reveals the coupling of spin with rotation \cite{14}-\cite{26}. As discussed in detail elsewhere~\cite{18}, for a positive-energy incident wave the energy as measured by an accelerated observer could be negative; nevertheless, this possibility cannot be ruled out as it is a reflection of the absolute character of the acceleration of the observer. Another possibility is that $E'_+$ could be zero. That is, by a mere rotation of frequency $\Omega =2E$, the rotating observers would perceive that the incident positive-helicity wave stands completely still since $E'_+=0$. This consequence of the hypothesis of locality is significantly modified by the nonlocal theory of accelerated systems.

The nonlocal ansatz implies, in the general case of uniformly rotating observers under consideration here, that
\begin{equation}\label{eq:41} \widehat{\Psi} _{\pm} =F_\pm (\tau )\widehat{\psi}_\pm \end{equation}
where $F_\pm$ is given by
\begin{equation}\label{eq:42} F_\pm (\tau )=1\pm \frac{1}{2}\gamma \Omega \frac{1-e^{iE'_\pm \tau }}{E'_\pm } .\end{equation}
It is interesting to note that for the observers under consideration here, Eq.~\eqref{eq:42} is of the general form~\cite{5,11}
\begin{equation}\label{eq:43} ^sF_\pm (\tau )=\frac{E\mp s\Omega e^{i{^sE}'_\pm \tau }}{E\mp s\Omega }\end{equation}
with $^sE'_\pm=\gamma (E\mp s\Omega )$ for $s=1/2$. As $^sE'_+\mapsto 0$, the incident wave does not stand still since $^sF_+\mapsto 1-is\gamma \Omega \tau$. This linear divergence in time, which is characteristic of resonance, can be avoided with a finite incident wave packet. Moreover, $^sF_-\mapsto \cos (s\gamma \Omega \tau ) \exp (is\gamma\Omega \tau )$ for $E\mapsto s\Omega$. Let us note here another direct consequence of nonlocality that is evident in Eq.~\eqref{eq:43}: The amplitude of a positive-helicity spinor of energy $E>s\Omega$ is enhanced by $E/(E-s\Omega )$, while that of the corresponding negative-helicity spinor is diminished by $E/(E+s\Omega )$. Thus the measured amplitude is enhanced (diminished) if the observer rotates about the direction of incidence in the same (opposite) sense as the helicity of the incident particle.

We now turn to the derivation of the nonlocal Dirac equation for these accelerated observers.

\section{Special class of rotating observers\label{sec:4}}

Consider the uniformly rotating observers in Sec.~\ref{sec:3} that remain fixed at $r=0$ all along the $z$ axis. These special observers remain fixed in space and refer their measurements to the standard axes of the global inertial frame for $t<0$, while for $t\geq 0$ they are ``uniformly rotating" as they refer their measurements to uniformly rotating axes. Suppose that at each event in a global inertial frame there is such an observer; at any given time $t$, the observers are identical except for the fact that each occupies a different fixed position in space. We note that this class of observers (for $t\geq 0$) is essentially the one considered in Ref.~\cite{13}. For these special observers, $k$ and $\Lambda$ can be simply obtained from Eqs.~\eqref{eq:32}-\eqref{eq:35} with $\beta =0$ and $\gamma =1$.

It is a general consequence of the Volterra integral equation
\begin{align}\label{eq:44} \widehat{\Psi} (\tau )&=\widehat{\psi}(\tau )+\int^\tau _{\tau _0} k(\tau ')\widehat{\psi} (\tau ')d\tau '\\
\intertext{that}
\label{eq:45} \widehat{\psi}(\tau ) &=\widehat{\Psi} (\tau )+\int^\tau _{\tau_0} r(\tau ,\tau ') \widehat{\Psi} (\tau ')d\tau ',\end{align}
where $r(\tau ,\tau ')$ is the resolvent kernel discussed in detail in Ref.~\cite{27}. Let us define the nonlocal Dirac spinor $\Psi$ along the worldlines of the accelerated observers such that
\begin{equation}\label{eq:46} \widehat{\Psi}=\Lambda \Psi.\end{equation}
It follows that
\begin{equation}\label{eq:47} \psi (\tau )=\Psi (\tau )+\int^\tau _{\tau _0} \tilde{r}(\tau ,\tau ')\Psi (\tau ')d\tau ',\end{equation}
where $\tilde{r}$ is related to the resolvent kernel by
\begin{equation}\label{eq:48} \tilde{r} (\tau ,\tau')=\Lambda^{-1}(\tau )r(\tau ,\tau ')\Lambda (\tau ').\end{equation}
The Lorentz-invariant nonlocal Dirac equation can be derived using Eq.~\eqref{eq:47}, since $\psi (x)$ satisfies Dirac's equation. The general nonlocal ansatz is manifestly invariant under Poincar\'e transformations of the background spacetime; therefore, the resulting nonlocal wave equation for $\Psi$ has the same symmetry.

For the special class of uniformly rotating observers, the results of Appendix~\ref{sec:A} indicate that $\tilde{r}=-k$. Thus Eq.~\eqref{eq:47} may be written for the observer fixed at $\mathbf{x}$ as
\begin{equation}\label{eq:49} \psi (t,\mathbf{x})=\Psi (t,\mathbf{x}) -u(t)k\int^t_0 \Psi (t',\mathbf{x})dt',\end{equation}
where $k=-(i\Omega /2)\sigma ^{(1)(2)}$ and $u(t)$ is the unit step function such that $u(t)=1$ for $t>0$ and $u(t)=0$ for $t<0$. Let us write Dirac's equation in the form
\begin{equation}\label{eq:50} i\frac{\partial \psi}{\partial t}=H\psi,\end{equation}
where
\begin{equation}\label{eq:51} H=\gamma^{(0)} [-i\gamma ^{(j)}\partial _j+m]\end{equation}
is the free Dirac Hamiltonian. The nonlocal Dirac equation can thus be written as
\begin{equation}\label{eq:52} i\frac{\partial \Psi}{\partial t}=H\Psi +iu(t)\left[ k\Psi +iHk\int^t_0\Psi (t',\mathbf{x})dt'\right]\end{equation}
for the special class of uniformly rotating observers that are fixed in space. A more complicated form of the nonlocal equation emerges for linearly accelerated observers discussed in the next section.

\section{Linearly accelerated observers\label{sec:5}}

Let us suppose that the fundamental inertial observers, at rest for $-\infty <t<0$, start from rest at $t=\tau =0$ and accelerate along the $z$ axis with acceleration $g(\tau )\geq 0$. Thus each such observer carries a tetrad frame for $t\geq 0$ given by
\begin{align}\label{eq:53} \lambda^\mu _{\;\;(0)} &=(C,0,0,S), & \lambda ^\mu_{\;\;(1)}&=(0,1,0,0),\\
\label{eq:54} \lambda^\mu_{\;\;(2)}&=(0,0,1,0), & \lambda^\mu _{\;\;(3)} &=(S,0,0,C).\end{align}
Here $C=\cosh \theta$, $S=\sinh \theta$ and
\begin{equation}\label{eq:55} \theta =\int^\tau _0 g(\tau ')d\tau '.\end{equation}
It follows from Eqs.~\eqref{eq:26} and \eqref{eq:27} that
\begin{align}\label{eq:56} k&=-\frac{i}{2}g(\tau )\sigma ^{(0)(3)},\\
\label{eq:57} \Lambda &=\cosh \frac{\theta}{2}+i\sinh \frac{\theta}{2}\;\sigma^{(0)(3)},\end{align}
since in this case $\Lambda_0$ is the identity matrix. It is shown in Appendix~\ref{sec:A} that $\tilde{r}(\tau ,\tau ')=-k(\tau ')$ in this case. Thus Eq.~\eqref{eq:47} may be written as
\begin{equation}\label{eq:58} \psi (\tau )=\Psi (\tau )-u(\tau )\int^\tau_0 k(\tau ')\Psi (\tau ')d\tau '.\end{equation}

One can use these results to study the extent of, and the influence of nonlocality on , the spin-acceleration coupling for a Dirac particle with $m>0$. This issue has a long history and requires a detailed investigation, which is beyond the scope of this paper~\cite{28}-\cite{30}.

To simplify matters, we assume that $g(\tau )$ vanishes except in the interval $(0,\tau_f)$, where it is a constant $g_0$. That is,
\begin{equation}\label{eq:59} g(\tau )=g_0[u(\tau )-u(\tau -\tau _f)].\end{equation}
This means that the class of accelerated observers undergoes hyperbolic motion from $t=0$ to $t_f$; then, for $t>t_f$ the observers move uniformly with speed $\beta_f$. Thus for $0\leq t\leq t_f$, $(0,x_0,y_0,z_0)\mapsto (t,x,y,z)$ along the worldline of an observer such that $x=x_0$, $y=y_0$,
\begin{equation}\label{eq:60} z=z_0+\frac{1}{g_0}(-1+\cosh g_0\tau ),\quad t=\frac{1}{g_0}\sinh g_0\tau .\end{equation}
Hyperbolic motion ends at $t_f$ such that $g_0t_f=\sinh g_0\tau _f$ and $z_f=z_0+(-1+\cosh g_0\tau _f)/g_0$; for $t>t_f$, the motion is uniform with speed $\beta _f=\tanh g_0\tau _f$.

It proves useful to define $\zeta (t)$ by
\begin{equation}\label{eq:61} \zeta (t)=\sqrt{t^2+\frac{1}{g^2_0}}\end{equation}
such that during hyperbolic motion $z-\zeta$ is a constant; that is, $z=z_0-g^{-1}_0+\zeta (t)$. Then Eq.~\eqref{eq:58} can be expressed as
\begin{equation}\label{eq:62} \psi (t,x,y,z)=\Psi (t,x,y,z)-\frac{i}{2}u(t)\sigma ^{(0)(3)}\mathcal{I},\end{equation}
where
\begin{equation}\label{eq:63} \mathcal{I}=\int^t_0 \Psi (t',x,y,z-\zeta +\zeta ')\frac{\mathcal{U}(t')}{\zeta '} dt'.\end{equation}
Here $\zeta '=\zeta (t')$, $z(t')=z-\zeta +\zeta '$ and
\begin{equation}\label{eq:64} \mathcal{U}(t)=u(t)-u(t-t_f).\end{equation}
Equation~\eqref{eq:62} is in a form that is suitable for the derivation of the nonlocal Dirac equation in this case. That is, applying the Dirac operator to Eq.~\eqref{eq:62} results in
\begin{equation}\label{eq:65} [i\gamma^{(\alpha )}\partial_\alpha -m]\Psi =\Delta ,\end{equation}
where $\Delta $ can be expressed as
\begin{equation}\begin{split} \label{eq:66} \Delta &=-\frac{i}{2}\frac{\mathcal{U}}{\zeta} \gamma^{(3)}\Psi -\frac{i}{2} mu\sigma ^{(0)(3)} \mathcal{I}-\frac{i}{2} u\left[\gamma^{(0)}-\frac{t}{\zeta} \gamma^{(3)} \right] \mathcal{I}_3\\
&\quad -\frac{1}{2}u\gamma^{(0)}\sigma ^{(3)(i)}\mathcal{I}_i .\end{split}\end{equation}
Here
\begin{equation}\label{eq:67} \mathcal{I}_i =\int^t_0 \Psi_{,i} (t',x,y,z-\zeta+\zeta ')\frac{\mathcal{U}(t')}{\zeta '}dt'\end{equation}
and $\Psi _{,i}=\partial _i \Psi$.

It is important to note that as $g_0\to 0$, $\zeta \to \infty$ and $\Delta \to 0$. Moreover, for $t>t_f$, $\Delta$ is in general nonzero; therefore, Dirac's equation is in general nonlocal for observers that are in uniform motion but have a history of past acceleration.

\section{Discussion\label{sec:6}}

The nonlocal determination of Dirac spinors by accelerated observers has been explored in this paper and the effect of nonlocality
on the phenomenon of spin-rotation coupling for a Dirac particle has been calculated. The Lorentz-invariant nonlocal Dirac equation
has been discussed and explicitly worked out for certain special classes of accelerated systems. It has been shown that, as a result
of the memory of past acceleration, nonlocality in general persists even after the observer's acceleration has been turned off.

It is essential that a proper nonlocal extension of the standard relativity theory of accelerated systems be consistent with the
quantum theory~\cite{5,31}. This appears to be the case for the motion of a {\it free} Dirac particle from the standpoint of an
accelerated observer in Minkowski spacetime considered in this paper. It is necessary to develop the theory further in order to
take interactions into account; however, this remains a task for the future.

\section*{ACKNOWLEDGEMENTS}

I am grateful to Friedrich Hehl for valuable discussions.

\appendix

\section{Resolvent Kernel\label{sec:A}}

The resolvent kernel $r$ can in general be calculated using a certain sum of iterated kernels~\cite{27}; in particular
\begin{equation}\label{eq:A1} r(\tau ,\tau ')=\sum^\infty_{n=1}(-1)^n k_n (\tau ,\tau '),\end{equation}
where
\begin{equation}\label{eq:A2} k_1 (\tau ,\tau ')=k(\tau ')\end{equation}
and
\begin{equation}\label{eq:A3} k_{n+1} (\tau ,\tau ')=\int^\tau_{\tau '}k(t)k_n (t,\tau ')dt\end{equation}
for $n=1,2,3,\dots $. The derivation and properties of resolvent kernels $r$ and $\tilde{r}$ are described in detail in Appendix A of Ref.~\cite{27}.

Let us first suppose that $k(\tau )$ is a constant matrix as in the case of uniformly rotating observers. Then it is possible to show that $\tilde{r}(\tau ,\tau ')$ is constant as well. It is given explicitly by
\begin{equation}\label{eq:A4} \tilde{r}=-\Lambda ^{-1}(\tau _0)k\Lambda (\tau _0),\end{equation}
as shown in detail in Appendix C of Ref.~\cite{27}. For the special class of rotating observers in Sec.~\ref{sec:4}, $\Lambda (\tau _0)=1$ for $\tau _0=0$ and hence $\tilde{r}=-k$.

We now turn to the case of linear acceleration in Sec.~\ref{sec:5}, where
\begin{equation}\label{eq:A5} k(\tau ')=-\frac{i}{2}g(\tau ')\sigma ^{(0)(3)}.\end{equation}
We find from Eqs.~\eqref{eq:A2} and \eqref{eq:A3} that
\begin{equation}\label{eq:A6} k_{n+1} (\tau ,\tau ')=\frac{\delta^n}{2^{n+1} n!} g(\tau ')M_n,\end{equation}
where $n=0,1,2,\dots $, and $\delta $ is defined by
\begin{equation}\label{eq:A7} \delta =\theta (\tau )-\theta (\tau ').\end{equation}
Here $M_n$ is a constant matrix given by
\begin{equation}\label{eq:A8} M_{2n}=-i\sigma^{(0)(3)},\quad M_{2n+1}=1.\end{equation}
It follows from Eq.~\eqref{eq:A1} that
\begin{equation}\label{eq:A9} r(\tau ,\tau ')=\frac{1}{2}g(\tau ')\left[ \sinh \frac{\delta}{2}+i\cosh \frac{\delta}{2}\;\sigma^{(0)(3)}\right],\end{equation}
where $\delta$ is given by Eq.~\eqref{eq:A7}. To find $\tilde{r}(\tau ,\tau ')$, we recall that
\begin{align}\label{eq:A10} \Lambda (\tau )&=\cosh \frac{\theta (\tau )}{2}+i\sinh \frac{\theta (\tau )}{2}\;\sigma^{(0)(3)},\\
\label{eq:A11} \Lambda^{-1}(\tau )&=\cosh \frac{\theta (\tau )}{2}-i\sinh \frac{\theta (\tau)}{2}\;\sigma ^{(0)(3)}.\end{align}
Calculating $\tilde{r}$ on the basis of its definition given in Eq.~\eqref{eq:48}, we find using Eqs.~\eqref{eq:A9}-\eqref{eq:A11} that $\tilde{r}(\tau ,\tau ')=-k(\tau ')$ in this case as well and hence
\begin{equation}\label{eq:A12} \tilde{r}(\tau ,\tau ')=\frac{i}{2} g(\tau ')\sigma^{(0)(3)},\end{equation}
which leads to Eq.~\eqref{eq:58}.

\end{document}